\documentclass[preprint,sort&compress]{elsarticle}
\usepackage{lineno,hyperref}
\modulolinenumbers[200]
\usepackage{revsymb4-1}
\usepackage{bm}
\usepackage{mathptmx}
\usepackage{braket}
\usepackage{bm}
\journal{Journal of \LaTeX\ Templates}

\begin{document}

\begin{frontmatter}
\title{Quadrupole Interaction of Non-diffracting Beams with Two-Level Atoms}

\author[mymainaddress]{Saud Al-Awfi}
\ead{alawfi99@hotmail.com}

\author[mysecondaryaddress]{Smail Bougouffa\corref{mycorrespondingauthor}}
\cortext[mycorrespondingauthor]{Corresponding author}
\ead{sbougouffa@hotmail.com and sbougouffa@imamu.edu.sa}

\address[mymainaddress]{Department of Physics, Faculty of Science, Taibah University,
P.O.Box 30002, Madinah, Saudi Arabia}
\address[mysecondaryaddress]{Physics Department, College of Science, Al Imam Mohammad ibn Saud Islamic University (IMSIU), P.O. Box 90950, Riyadh 11623, Saudi Arabia.}

\begin{abstract}
Recently it has been shown that the quadrupole interactions can be improved significantly as the atom interacts at near resonance with the Laguerre-Gaussian (LG) mode. In this paper, we illustrate that other kinds of optical vortex can be also led to a considerable enhancement of quadrupole interaction when the atom interacts with optical modes at near resonance. The calculations are performed on an interesting situation with Cs atom, where the process is concerned with dipole-forbidden and quadrupole-allowed transitions with a convenable choice of atomic and optical mode parameters.  In this direction,  we show that the quadrupole transitions can be significantly enhanced and therefore they can play an interesting role and lead to new features of atom-light interaction, which can have some constructive implications in experiments.
\end{abstract}

\begin{keyword}
\texttt{Quadrupole interaction; non-diffracting beams; atom-field interactions; Optical forces; Optical Vortices.}
\end{keyword}

\end{frontmatter}

\linenumbers

\section{Introduction}\label{sec1}
Since the emergence of atom optics during last decades, both theoretical and experimental investigations have been focused on the active-dipole interaction. This center of attention is reasonable and is mainly due to the control of this kind of interaction on the outcome situation between the atom and light field \cite{loudon2000quantum}.

Furthermore, most of the models in quantum optics consider only the electric dipole interaction between systems and the electromagnetic field. In particular, the interaction of nearly resonant laser beams with atoms, which can be formulated as two-level system interacting with laser light where the dipole active transitions are involved \cite{allen1987optical,grynberg2010introduction}. In the context of dipole approximation, the studies of the aspects of diffraction of atoms in a laser field have guided to a several useful application, including laser cooling, Bose-Einstein condensation, ultra cold atoms, atom lasers, simulation of condensed matter systems, production and study of strongly correlated systems, and production of ultra cold molecules \cite{letokhov2007laser,claude2011advances,haroche2006exploring}.

The useful experimental progressions on the photon processes concerning the interaction of atoms and molecules with lasers, suggest deep theoretical investigations on atom-light interaction. In addition, recent advances in the optical measurement techniques on quadrupole transitions have been realized and employed \cite{tojo2004absorption,kern2011excitation,cheng2012cavity}. Then, some appropriate theoretical examinations that are concerned with the exploration of the quadrupole interaction effects have been suggested \cite{klimov1996quadrupole,kern2012strong,lembessis2013enhanced}. Some such studies are concerned with the exploration of the effects of magnetic-dipole interaction, in the context of the magnetic mirrors applications \cite{choi2015near,lin2016dielectric,liu2017generalized}, while others are dealt with the electric-quadrupole aspects in the framework of the atoms motion in a laser field \cite{tojo2004absorption,kern2011excitation,cheng2012cavity,klimov1996quadrupole,kern2012strong,lembessis2013enhanced} and other studies are concerned with the chiroptical interactions using vortex light, which can only occur through the quadrupole transitions \cite{forbes2018optical,forbes2018spin}.

On the other hand, optical vortices can be considered as one of the most attractive branch that has emerged in atom optics. In the beginning their realization requires the cylindrical optical lenses \cite{beijersbergen1993astigmatic}. However, with the ability of the experimental generation of LG beams, the optical vortices can easily be created in space regions \cite{barnett2016optical,andrews2011structured}. More advances in the vortex laser beams that are characterized by a quantized orbital angular momentum, are established experimentally and determined through the surfaces optical vortices \cite{lembessis2011surface}, which lead to an actual presence of optical tweezers \cite{novotny1997theory,moffitt2008recent,neuman2008single}. 

In fact, the two main types of light that have considered in this area were LG  and Bessel  beams. Both beams are illustrated by the property of orbital angular momentum for all light modes greater than the fundamental mode \cite{al2012generation,lembessis2009surface}. For both beams, the fundamental mode is only able to produce the optical lattices \cite{al2010optical}. This interesting characteristic is the main feature of Hermite-Gaussian light, which leads us to investigate Hermite-Gaussian light in this study beside the light types possessing orbital angular momentum. However, the feature of orbital angular momentum that possesses the two first types of beams makes them of particular significance in the case of the torque that robustly influences the rotational atomic motion.

The major goal of this paper is to extend the explorations on  quadrupole interactions in order to exploit the optical forces that are responsible on the motion and trapping small objects, neutral atoms, ions and molecules to well defined space regions. We will illustrate that the inclusion of the quadrupole interactions for some specific cases with forbidden dipole transitions can lead to interesting aspects. In particular, some recent investigations have pointed out the possibility of observing quadrupole interaction and even enhancing its magnitude \cite{forbes2018spin,forbes2018optical,filter2012controlling,lembessis2013enhanced}. Moreover, the realization of optical lattices and vortices of atoms in space, in particular, requires the use of special types of light that are described as possessing non-diffracting beams. These kinds of beams can actually be produced in the laboratories at different orders.

Finally, the investigation of the quadrupole interaction effects for these three beams independently, gives an objective comparison between various aspects that can be generated with the different beams. Nevertheless, we can determine the most appropriate situation in the studies of laser light action on atoms.

The paper is organized as follows. In section \ref{sec2}, the quadrupole interaction formalism is presented for atom-field interactions. Section \ref{sec3} considers the mechanical motion of two level atom in the quadrupole interactions and the corresponding optical forces are identified. In section \ref{sec3}, we are concerned with different kind of optical beams and the expression of Rabi frequency is obtained in the quadrupole interaction. The spatial distribution of Rabi frequency is presented with typical parameters. Section \ref{sec4} provides comments and conclusions.

\section{Quadrupole interaction}\label{sec2}
The goal is to investigate the spatial distribution of the quadrupole Rabi frequency and its magnitude comparative to the case with different light modes. In addition, the effects of the optical vortices on this spatial distribution will be discussed.  Thus, the consequences of this quadrupole Rabi frequency on the atom dynamics due to the forces and torque acting on the atom in the field of the optical vortex will be explored.

We consider here a two-level atom interacting with an optical vortex propagating along the $z$ axis with an axial wave vector $k$. The interaction Hamiltonian in a multipolar series about the center of mass $\mathbf{R}$ can be read as
\begin{equation}\label{1}
    \hat{H}_{int}=\hat{H}_{dp}+\hat{H}_{qp}+...,
\end{equation}
where
\begin{equation}\label{2}
      \hat{H}_{dp}=- \hat{\bm{\mu}}.\mathbf{ \hat{E}}(\mathbf{R}),
\end{equation}
which represents the electric dipole interaction between system and the EM field, where $\bm{ \hat{\mu}}=q\mathbf{r}$ is the diploe moment and $\bm{ \hat{E}}(\mathbf{R})$ is the EM field.
The quadrupole term is explicitly defined as
\begin{equation}\label{3}
  \hat{H}_{qp}=-\frac{1}{2}\sum_{ij} \hat{Q}_{ij}\frac{\partial  \hat{E_j}}{\partial r_i},
\end{equation}
which represents the interaction between the quadrupole moments $\mathbf{ \hat{Q}}$ and the gradient of the electric field that is a function of the center-of-mass coordinate $\mathbf{R}$ and $r_i$ are the components of the internal position vector $\mathbf{r}$.
In order to simplify the problem, without loss of generality, we consider the electric field polarized along the $x$ direction, which yields the following form of the quadrupole interaction Hamiltonian
\begin{equation}\label{4}
    \hat{H}_{qp}=-\frac{1}{2}\Big(\hat{Q}_{xx}\frac{\partial  \hat{E}_x}{\partial X}+\hat{Q}_{xy}\frac{\partial  \hat{E}_x}{\partial Y}+\hat{Q}_{xz}\frac{\partial  \hat{E}_x}{\partial Z} \Big),
\end{equation}
where the elements of the quadrupole tensor operator for two-level system can be read as
\begin{equation}\label{5}
   \hat{Q}_{ij}=Q_{ij}( \hat{\pi} + \hat{\pi}^{\dag}),
\end{equation}
where $Q_{ij}=\bra{i}\hat{Q}_{ij}\ket{j}$ are the quadrupole matrix element and $ \hat{\pi} ( \hat{\pi}^{\dag})$ are the atomic level lowering and raising operators.

We assume that the EM field is linearly polarized along the $x$ direction such that its quantized electric field as a function of the center-of-mass in cylindrical coordinates $\mathbf{R}=(r,\phi, Z)$ can be read as
\begin{equation}\label{6}
    \mathbf{ \hat{E}}(\mathbf{R})=\mathbf{ \hat{i}} u_{\{k\}}(r)\hat{a}_{\{k\}}e^{i \Theta_{\{k\}}(\mathbf{R})}+H.c.
\end{equation}
where  $u_k(r)$ and  $\Theta_k(Z)$ are the amplitude and the phase of the EM field of the optical modes $\{k\}$. Here $\{k\}$ represents a group of indices that characterize the optical modes. $\hat{a}_{\{k\}}$ is the annihilation operator for the field mode and $H.c.$ refers to Hermitian conjugate.
By substituting this form for the EM field in Eq.(\ref{4}), we get the following form for the quadrupole interaction Hamiltonian
\begin{equation}\label{7}
    \hat{H}_{qp}=\hbar\hat{a}_{\{k\}}\Omega^{qp}_{\{k\}}(\mathbf{R})e^{i\Theta_{\{k\}}(\mathbf{R})}+H.c.
\end{equation}
where $\Omega^{qp}_{\{k\}}(\mathbf{R})$ is the complex Rabi frequency.

\section{Optical forces}\label{sec3}
Here, we link the previous formalism to the concept of mechanical force that acts by the EM field on the atom. Using the density matrix techniques, we can show that the steady state of total average quadrupole force $\left\langle F_{k}^{opt.} \right\rangle $ on the moving atom with velocity $\mathbf{V}=\dot{\mathbf{R}}$  is given by
\begin{equation}\label{8}
    \left\langle F_{k}^{opt.} (\mathbf{R},\mathbf{V})\right\rangle =\left\langle F_{k}^{spon} (\mathbf{R},\mathbf{V})\right\rangle +\left\langle F_{k}^{Q} (\mathbf{R},\mathbf{V})\right\rangle
\end{equation}
where
\begin{equation}\label{9}
   \left\langle F_{k}^{spon} \right\rangle =2\hbar \Gamma _{Q} \left|\Omega _{k}^{Q} (\mathbf{R})\right|^{2} \left({\nabla \theta _{k} (\mathbf{R})\over \Delta _{k}^{2} (\mathbf{R},\mathbf{V})+2\left|\Omega _{k}^{Q} (\mathbf{R})\right|^{2} +\Gamma _{Q}^{2} } \right),
\end{equation}
is the spontaneous force that represents the force due to the quadrupole absorption and re-emission of light by the atom and
\begin{equation}\label{10}
   \left\langle F_{k}^{Q} \right\rangle =-\hbar \nabla \left|\Omega _{k}^{Q} (\mathbf{R})\right|^{2} \left({\Delta _{k} (\mathbf{R},\mathbf{V})\over \Delta _{k}^{2} (\mathbf{R},\mathbf{V})+2\left|\Omega _{k}^{Q} (\mathbf{R})\right|^{2} +\Gamma _{Q}^{2} } \right),
\end{equation}
is the quadrupole force that arises from the nonconformity of the field distribution.  $\nabla \theta _{k} (\mathbf{R})$ stands for the gradient of the phase $\theta _{k} (\mathbf{R})$. $\Gamma _{Q} $  is the decay rate for the quadrupole spontaneous emission and $\Delta _{k} (\mathbf{R},\mathbf{V})$ is the dynamic detuning which is a function of both the position and the velocity vectors of the atom
\begin{equation}\label{11}
    \Delta _{k} (\mathbf{R},\mathbf{V})=\Delta _{0} -\mathbf{V}\cdot \nabla \theta _{k} (\mathbf{R}),
\end{equation}
where $\Delta _{0} =\omega -\omega _{0} $ is the static detuning, with $\hbar \omega _{0} $ is the atomic separation energy and $\omega $ is the light frequency. The second term in Eq. (\ref{11}) $\delta =-\mathbf{V}\cdot \nabla \theta _{k} (\mathbf{R})$ represents the Doppler shift due to the excited mode. Thus, the quadrupole force is responsible for confining the atom to maximal or minimal intensity regions of the field, depending on the detuning $\Delta _{k}$ . In fact, in a dipole-allowed transition the atom involves with the optical field strength whereas in a quadrupole transition it participates only with the field gradient. The gradients of the electric field in atom-field interactions can lead to transitions for atoms confined in the dark regions of the light beam where there is weak light intensity but relatively strong field gradients \cite{babiker2018atoms}.

The quadrupole force is derivable from a quadrupole potential $\left\langle F_{k}^{Q} \right\rangle =-\nabla \left\langle U_{k}^{Q} (\mathbf{R})\right\rangle $ with $U_{k}^{Q} (\mathbf{R})$ having the well known form

\begin{equation}\label{12}
    \left\langle U_{k}^{Q} (\mathbf{R})\right\rangle =-{\hbar \Delta _{0} \over 2} \ln \left(1+{2\left|\Omega _{k}^{Q} (\mathbf{R})\right|^{2} \over \Delta _{0}^{2} +\Gamma _{Q}^{2} } \right).
\end{equation}
The conventional derivation of average quadrupole force corresponding to Eq. (\ref{12}) makes use of density matrix methods leading to the average forces given by equation (8) emerging in the steady state. For red detuned light $\Delta _{0} <0$, the quadrupole potential exhibits a minimum in the high intensity region of the beam tuned below resonance where $\Delta _{0} <0$. For blue detuned $\Delta _{0} >0$, we have trapping in the low-intensity (dark) regions of the field. On the other hand, in many experiment situations the large detuning is considered $(\Delta _{0}\gg \mid\Omega^Q\mid)$ and $(\Delta _{0}\gg \Gamma_Q)$, then the quadrupole potential can be approximate by
\begin{equation}\label{13}
    \left\langle U_{k}^{Q} (\mathbf{R})\right\rangle \approx -\frac{\hbar}{\Delta _{0}} \left|\Omega _{k}^{Q} (\mathbf{R})\right|^2.
\end{equation}
Thus, from the previous brief formalism we can see that the modulus squared Rabi frequency plays a crucial role to explore the dynamics atom relative to the field of mode $k$.
\section{Optical Vortices}
In the following, we will illustrate the previous concepts with different kind of the optical vortices and investigate the quadrupole spatial distribution of Rabi frequency with the most important optical vortices that can be experimentally established and have large applications.
\subsection{Laguerre Gaussian Modes}
\begin{figure}[h]
\resizebox{1\linewidth}{!}{%
 \includegraphics{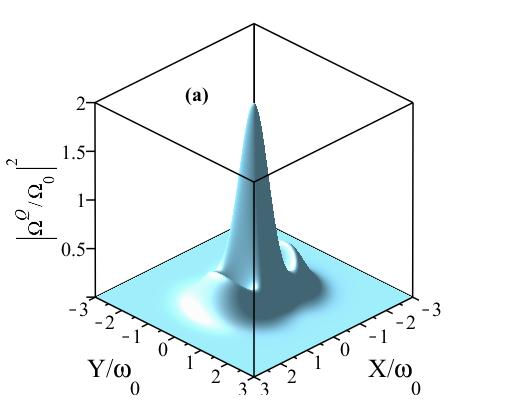}~\includegraphics{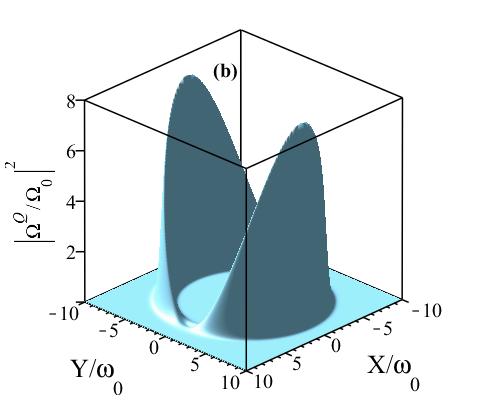}~\includegraphics{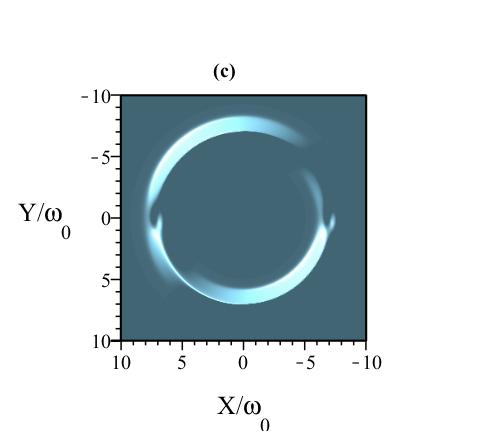}}
\caption{(Color online) The spatial distribution of the modulus of squared relative Rabi frequency $\left|\Omega_{klp}^Q/\Omega_0\right|^2$ for an atom in a (LG) doughnut mode. (a) in the case $\ell =1$ and $p=0$, (b) for the case $\ell=100$ and $p=1$. (c) The projection in the $X-Y$ plane of the case $\ell=100$ and $p=1$.}\label{Fig1}
\end{figure}

Recently, it has shown that the weak optical quadrupole interaction in atoms can be improved considerably while the atom interacts at near resonance with an optical vortex \cite{lembessis2013enhanced,forbes2018spin}. In addition, the LG mode is investigated in the case of the quadrupole approximation and with an appropriate choice of the winding number $l$ of the vortex, the process involving the dipole-forbidden, but quadrupole-allowed, transitions in atoms can be realized \cite{lembessis2013enhanced}. Indeed, from equation (\ref{6}) in the cylindrical coordinates , the LG mode is characterized by its amplitude and phase \cite{deng2008propagation,deng2010dynamics,deng2008hermite,babiker2018atoms}, which are given by
\begin{equation}\label{14}
    u_{\{k\}}(r)=u_p^{|\ell|}(r)=E_{k00}\sqrt{\frac{p!}{(|l|+p)!}}\Big( \frac{r\sqrt{2}}{\omega_0}\Big)^{|l|}L_p^{|l|}(\frac{2r^2}{\omega_0^2})e^{-r^2/\omega_0^2},
\end{equation}
where $L_p^{|l|}$ is the Laguerre polynomial of degree $p$ and $\omega_0$ is the radius at beam waist  at $Z=0$, $E_{k00}$ is the constant amplitude of the plane electromagnetic wave, and
\begin{equation}\label{15}
    \Theta_{klp}=skZ+l\phi -s(2p+|l|+1)\tan^{-1}(Z/z_{R})+s\frac{kr^2Z}{2(Z^2+z_{R}^2)},
\end{equation}
is the phase of the optical mode. In this equation, the third term is the Gouy phase for the LG mode and the fourth term represents the curvature phase. The parameter $s=\pm 1$ takes into account propagation in the opposite direction along the $\pm z$-axes. Nevertheless, this general phase form (\ref{15}) is needed for defining the various effects with rotational effects when the light interacts with atoms and molecules. It is frequently appropriate to emphasis on the form of twisted light without obvious details of the LG form.
 Then, we assume here that the LG mode has a large Rayleigh range $(z_R \rightarrow \infty)$. This situation is frequently faced and is achievable in practice. However, the previous phase can take the following form
\begin{equation}\label{15}
    \Theta_{klp}\approx kZ+l\phi,
\end{equation}
where we ignored all mode curvature effects, which frequently distract from the fundamental issues. However, the optical forces and trapping potentials due to the kind of vortex field were extensively explored \cite{lembessis2013enhanced,andrews2011structured,al2000atomic}. With LG mode, we get for the complex Rabi frequency $\Omega _{k\ell p}^{Q} $
\begin{equation}\label{16}
    \Omega _{k\ell p}^{Q} (\mathbf{R})=\left(u_{p}^{\ell } (r)/\hbar \right)\left (\alpha Q_{xx} +\beta Q_{yx} +ik Q_{zx} \right)
\end{equation}
where
\begin{eqnarray}
\alpha =\left(\frac{\left|\ell \right|X}{r^{2} } -\frac{2X}{w_{0}^{2} } -\frac{i\ell Y}{r^{2} } +\frac{1}{L_{p}^{\left|\ell \right|} } \frac{\partial L_{p}^{\left|\ell \right|} }{\partial X} \right) \label{17a}\label{17},\\
\beta =\left(\frac{\left|\ell \right|Y}{r^{2} } -\frac{2Y}{w_{0}^{2} } +\frac{i\ell X}{r^{2} } +\frac{1}{L_{p}^{\left|\ell \right|} } \frac{\partial L_{p}^{\left|\ell \right|} }{\partial Y} \right) \label{18}.
\end{eqnarray}
In order to illustrate the effect of the quadrupole effects with Laguerre-Gaussian (LG) mode, we limit our exploration to case that is recently considered in \cite{lembessis2013enhanced}. We consider the case of a LG doughnut mode of winding number $\ell$ and $p = 0$. In this case, the derivatives in $\alpha$ and $\beta$ in Eqs. (\ref{17},\ref{18}) are null, as $L_{0}^{\left|\ell \right|}$ is a constant for all $\ell$. In addition, we also suppose that the atom is constrained to move in the $X-Y$ plane and the quadrupole transition is such that $Q_{xy} =Q_{xz}=0$. Within these assumptions the Rabi frequency Eq. (\ref{16}) can be read as:
\begin{equation}\label{19}
    \Omega _{k\ell 0}^{Q} (\mathbf{R})=\left (u_{0}^{|\ell| }(r)/\hbar \right) Q_{xx}\Big( \frac{\left|\ell \right|X}{r^{2} } -\frac{2X}{w_{0}^{2} } -\frac{i\ell Y}{r^{2} }\Big).
\end{equation}
We perform the numerical calculations with an interesting case of the Cs atom, which is recently illustrated for its quadrupole transition $(6^2S_{1/2}\rightarrow5^2D_{5/2})$ with an appropriate value of the essential parameters $\omega_0=\lambda/2$, $\lambda=675 (nm)$, $Q_{xx}=10e a_0^2$, $\Gamma_Q=7.8\times 10^5 (s^{-1})$, $\Delta_0=10^3\Gamma_Q$ and for the intensity $I=\epsilon_0cE_{k00}^2/2=10^9Wm^{-2}$. In consequence, it is suitable to label the scaling factor of the Rabi frequency by $\Omega_0=\frac{1}{\hbar}\big(\frac{2I}{\epsilon_0 c}\big)^{1/2}\frac{Q_{xx}}{\omega_0}=136\Gamma_Q$. In Fig.\ref{Fig1}, we present the spatial distribution of $|\Omega_{k\ell 0}^Q|^2$ in units of $\Omega_0$ for doughnut vortex of winding numbers $|\ell|=1$ and $|\ell|=100$.\\

It is clear from Fig.\ref{Fig1}(a) that the presence of a maximum variation at the vortex core is maintained only of the case $\ell = 1$. While Fig.\ref{Fig1}(b) illustrates the analogous variation of the relative Rabi frequency for the large winding number $ l =100$. From the experiment point view, winding numbers as large as $l = 300$ can be accomplished, as shown recently \cite{fickler2012quantum}. On the other hand, the shape of the relative quadrupole Rabi frequency in the LG mode shows augmentation of both the trapping potential and the dissipative force in the procedures characterized by quadrupole transitions. These important consequences are recently explored \cite{lembessis2013enhanced} and pointed out that these results can provide significant mechanical effects on atoms categorized by a quadrupole-allowed transition, which must be considered to improve the experimental investigations.

\subsection{Bessel Modes}
In this section, we follow same previous procedure to explore the quadrupole interaction of non-diffracting Bessel beam, which is essentially characterized at general position  $R=(r,\varphi ,Z)$ in cylindrical polar coordinates, by its phase $\theta _{km} $ and the complex Rabi frequency $\Omega _{km}^{Q}$ that are given as \cite{al2010optical}
\begin{equation}\label{20}
    \theta _{km} (\varphi ,Z)=kZ+m\varphi,
\end{equation}
\begin{equation}\label{21}
    \Omega _{km}^{Q} (r,Z)=\left(g_{m}^{} (r)/\hbar \right)\left [Q_{xx} \eta +Q_{xy} \mu +Q_{xz} \sigma \right],
\end{equation}
where $\textbf{k}$ is the beam  axial wave vector and $m$ is the quantum number. The parameters  $\eta ,\mu $ and $\sigma $ are given respectively as
\begin{eqnarray}
  \eta (r)&=&\left({1\over J_{m} } {\partial J_{m} \over \partial X} -{imY\over r^{2} } \right)\label{22}, \\
 \mu (r)&=&\left({1\over J_{m} } {\partial J_{m} \over \partial Y} +{imX\over r^{2} } \right)\label{23}, \\
 \sigma (r,Z)&=&\left({(2m+1)\over 2Z} -{-2Z\over Z_{\max }^{2} } +ik_{Z} \right)\label{24}.
\end{eqnarray}
The Bessel amplitude function $g_{m}^{} (r)$ can be read as
\begin{equation}\label{25}
    g_{m}^{} (r)=\sqrt{{8\pi ^{2} k_{\bot }^{2} w_{0}^{2} I\over \varepsilon _{0} c} } \left({Z\over Z_{\max } } \right)^{m+1/2} \exp \left(-{2Z^{2} \over Z_{\max }^{2} } \right)J_{m} \left(k_{\bot } r\right),
\end{equation}
with $J_{m} $ denoting the $m^{th} $-order Bessel function of the first kind, $r^{2} =X^{2} +Y^{2} $, $k_{\bot }^{2} +k_{Z}^{2} =k_{0}^{2} $ and $k_{0}^{} =(2\pi /\lambda )$ being the wave number in free space while $I$ is the beam intensity. In addition, $k_{\bot } =k_{0} \sin \alpha $ and $k_{Z} =k_{0} \cos \alpha $ are the transversal and longitudinal components of the wave vector respectively, while $\alpha $ is the opening angle of the cone on which the wave traverses. In addition, $w_{0}^{} $ is the input Gaussian beam's waist, and $Z_{\max } $ is the typical ring spacing \cite{mcgloin2003three}.  We point out here that the central spot of the zero-order Bessel mode (denoted by $J_{0} $) is always bright (a central maximum) whereas that of all higher-order Bessel modes (denoted by $J_{m} $, $m$ being an integer and $\ge 1$) are always dark on the axis and are surrounded by concentric rings whose peak intensities decrease as $r^{-1} $ \cite{arlt2000generation}. Additionally, only the higher-order Bessel modes with $m\ge 1$ have an azimuthal phase dependence, $e^{im\varphi} $, on the mode axis, and, therefore, have a non-diffracting dark core. This property is directly related to the orbital angular momentum carried by the light modes, which is an addition to any spin angular momentum associated with their wave polarization. On the other hand, the factors $\hat{Q}_{ij} $ are the elements of the quadrupole tensor operator defined in Eq.(\ref{5}).\\
For the numerical calculations, it is instructive therefore to consider the previous important case of the Cs atom that is recently exemplified for its quadrupole transition $(6^2S_{1/2}\rightarrow5^2D_{5/2})$ with suitable values of the crucial parameters,
 which are illustrated previously. Lastly, we shall also assume that the atom is constrained to move in the $XY$ plane and the quadrupole transition is such that $Q_{xy} = Q_{xz}=0 $.\\

\begin{figure}[h]
\resizebox{1\linewidth}{!}{%
 \includegraphics{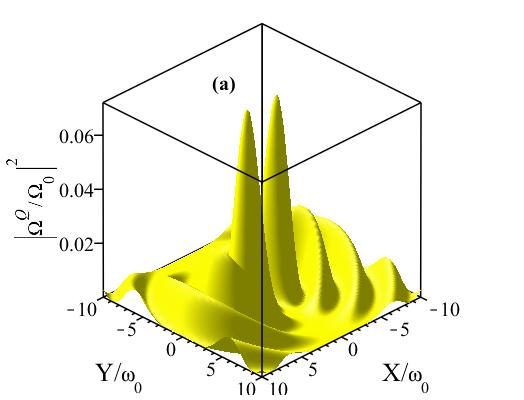}~\includegraphics{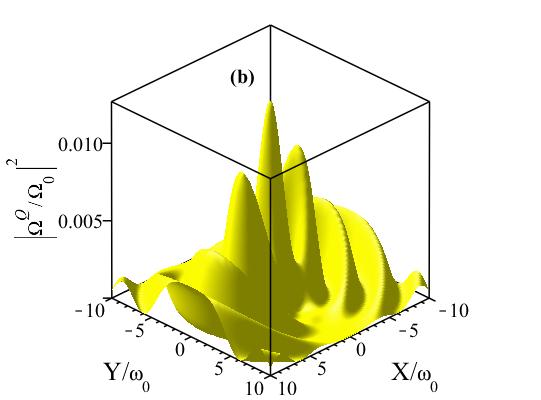}}\\
\resizebox{1\linewidth}{!}{%
 \includegraphics{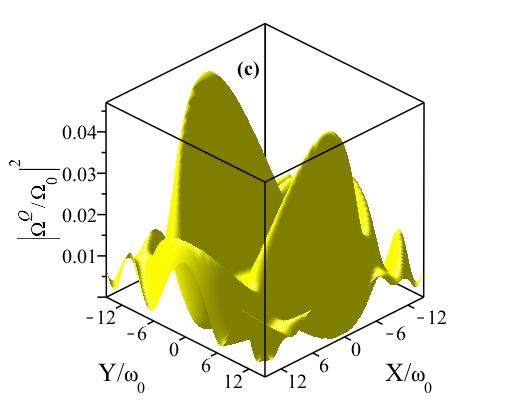}~\includegraphics{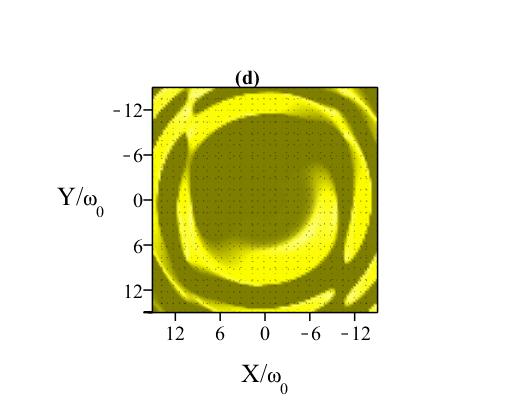}}
\caption{(Color online) The spatial distribution of the modulus of squared relative Rabi frequency $\left|\Omega_{klp}^Q/\Omega_0\right|^2$ for an atom in a Bessel mode for $Z=Z_{max}$. (a) in the case azimuthal mode $m =0$, (b) for the case $m=1$. (c) for the case $m=10$. (d) The projection in the $X-Y$ plane of the case $m=10$.  }\label{Fig2}
\end{figure}
Here , it is appropriate to label the scaling factor of the Rabi frequency by $\Omega_0=\frac{Q_{xx}}{\hbar \omega_0}\sqrt{{8\pi ^{2} k_{\bot }^{2} w_{0}^{2} I\over \varepsilon _{0} c} }$. In Fig.\ref{Fig2}, we point out  the spatial distribution of $|\Omega_{k m}^Q|^2$ in units of $\Omega_0$ for high-order vortex beam  $m=0$, $m=1$ and $m=10$.\\

We consider the rotational features including a rotational shift due to the azimuthal dependence of the field structure which arises in the interaction of the atom with any Bessel mode of order $m>0$. 

\subsection{Hermite-Gaussian Modes}
In the active-quadrupole interaction, the essential features of a Hermite-Gaussian beam, at a general position $R=(X,Y,Z)$, are the phase $\theta _{knm} (R)$ and the complex Rabi frequency $\Omega _{knm}^{Q} (R)$. These are written as [1]
\begin{equation}\label{26}
   \theta _{knm} (Z)=(n+m+1)\tan ^{-1} \left(Z/Z_{R} \right)+kZ,
\end{equation}
and
\begin{equation}\label{27}
    \Omega _{knm}^{Q} (R)={\xi _{k00} \over 2\hbar } \cdot C_{nm} \cdot F_{knm} (R)\left\{\alpha Q_{xx} +\beta Q_{xy} +\gamma Q_{xz} \right\},
\end{equation}
respectively, where $\textbf{k}$ is the axial wave vector of the beam mode,
where the factor $\xi _{k00} $ is the amplitude for a corresponding plane wave of
intensity $I$ propagating in the dielectric medium of refractive index $\eta$:

\begin{equation}\label{28}
    \xi _{k00} =\sqrt{2I/\eta ^{2} \varepsilon _{0} c}
\end{equation}
and $C_{nm} $ is the normalization constant of the Hermite-Gaussian function and is given by
\begin{equation}\label{29}
    C_{nm} =\left[2/\left(2^{n+m} _{} n!^{} m!^{} \pi \right)\right]^{1/2},
\end{equation}
for the case $n\ge m$,  while $F_{knm} (R)$ can be read as
\begin{eqnarray}\label{30}
    F_{knm} (R)&=&\frac{w_{0}}{ w(Z)} \exp \left[-ik{(X^{2} +Y^{2} )\over 2R(Z)} \right]\times \exp \left[{-(X^{2} +Y^{2} )\over w^{2} (Z)} \right] \nonumber \\ && \times H_{n} \left({\sqrt{2} X\over w(Z)} \right)\times H_{m} \left({\sqrt{2} Y\over w(Z)} \right).
\end{eqnarray}
Here $R(Z)=(Z_{R}^{2} +Z^{2} )/Z$ is the radius of curvature of the mode's wavefront with $Z_{R} $ the Rayleigh range; $w(Z)$ is the radius at which the Hermite-Gaussian mode amplitude and intensity drop to $1/e$ and $1/e^{2} $ of their axial values, respectively,
\begin{equation}\label{31}
    w^{2} (Z)=(Z_{R}^{2} +Z^{2} )/kZ_{R}.
\end{equation}
The position $Z=0$, referred to as the Hermite-Gaussian mode waist, corresponds to the waist size $w_{0} $ of the Hermite-Gaussian mode, such that: $w_{0}^{2} =2Z_{R} /k$. The special function $H_{n} (.)$ is the Hermite polynomial of order $n$.  For this reason, Hermite-Gaussian modes are typically designated $TEM_{nm} $ where {\it m} and {\it n} are the polynomial indices in the $X$ and $Y$ directions. Here $(X,Y)$ are transverse coordinates in the Cartesian coordinate systems. On the other hand, $\alpha ,\beta $ and $\gamma $  are given respectively by
\begin{eqnarray}
  \alpha &=&{1\over H_{n} } {\partial H_{n} \over \partial X} -ik{X\over R(Z)} -{2X\over w^{2} (Z)}\label{32}, \\
  \beta &=&{1\over H_{m} } {\partial H_{m} \over \partial Y} -ik{Y\over R(Z)} -{2Y\over w^{2} (Z)}\label{33}, \\
  \gamma &=&{1\over H_{n} } {\partial H_{n} \over \partial Z} +{1\over H_{m} } {\partial H_{m} \over \partial Z} -{1\over R(Z)} +ik+{i(n+m+1)\over kw^{2} (Z)} \nonumber \\ & &+{(X^{2} +Y^{2} )\over R(Z)} \left[{ik\over R(Z)} -{ik\over 2Z} +{2\over w^{2} (Z)} \right]\label{34}.
\end{eqnarray}
Assuming that the atom is constrained to move in the $XY$ plane and the quadrupole transition is such that  $Q_{xy} =Q_{xz}=0 $. Under this condition the electric-quadrupole interaction Eq.(\ref{27}) takes the following form:
\begin{equation}\label{34}
    \Omega _{knm}^{Q} (R)={\xi _{k00} \over 2\hbar } \cdot C_{nm} \cdot F_{knm} (R)\left\{Q_{xx} \left({1\over H_{n} } {\partial H_{n} \over \partial X} -ik{X\over R(Z)} -{2X\over w^{2} (Z)} \right)\right\}.
\end{equation}
In the following we will consider the same previous case for the Cs atom with the specific values of the essential parameters.
\begin{figure}[h]
\resizebox{1\linewidth}{!}{%
 \includegraphics{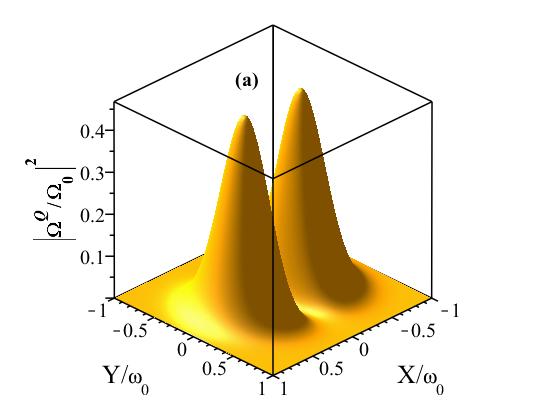}~\includegraphics{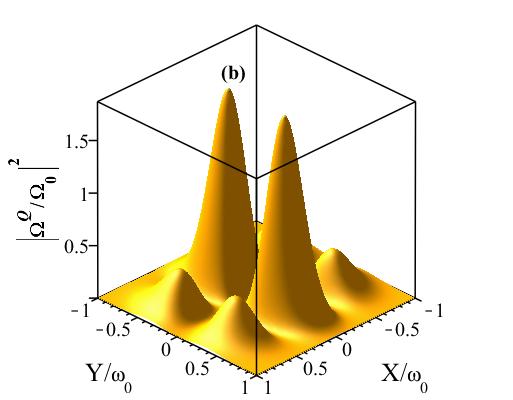}~\includegraphics{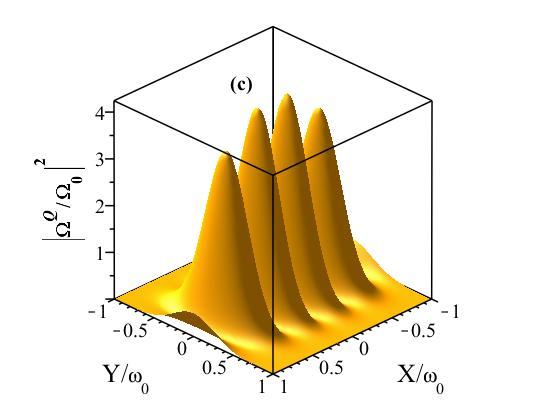}}
\caption{(Color online) The spatial distribution of the modulus of squared relative Rabi frequency $\left|\Omega_{klp}^Q/\Omega_0\right|^2$ for an atom in a Hermite Gaussian mode. (a) in the case $n=1$ and $m=0$, (b) for the case $n=1$ and $m=1$. (c) For the case $n=2$ and $m=0$.}\label{Fig3}
\end{figure}
In this case, we label the scaling factor of the Rabi frequency by $\Omega_0=\frac{Q_{xx}}{\hbar \omega_0}\sqrt{2I/\eta ^{2} \varepsilon _{0} c} $. In Fig.\ref{Fig3}, we show the spatial distribution of $|\Omega_{k m}^Q|^2$ in units of $\Omega_0$ for different order Hermite-Gaussian beam  $(n=1, m=0)$, $(n=1, m=1)$ and $(n=2, m=0)$.The spatial distribution of such  optical mode has a symmetric reparation  in the $XY-$plane with  $n$  maximum nodes in the horizontal direction and $m$ maximum nodes in the vertical direction.  followed by (2n-1). For $n=m = 0$, this case corresponds to the Gaussian beam  can be obtained Fig.\ref{Fig3}-a . This mode is called the fundamental mode or axial mode in the quadrupole regime and it has the highest beam quality. Other Hermite-Gaussian modes with indices $n$ and $m$ have an improved amplitude factor of $(2n)$ order in the x direction, and $(2m)$ in the y direction in the quadrupole interaction.

\section{Conclusions}\label{sec4}
In this work, we have presented the derivations of the quadrupole-active transition, acting on a two-level atom moving in three different optical beams. We have shown that the value of the Rabi frequency and hence the quadrupole trapping potential is sufficient enough to exploit in optical manipulation.

From previous experience in the active-dipole interaction studies we know that this result can be enhanced by several techniques such as a combination of strong field gradients and high field intensity or using the mutual coupling of two co-propagating and counter-propagating beams \cite{monden2012interaction,lai1997radiation}. The mutual technique, in particular assists to avoid some undesirable effects of single beam interaction such as destabilizing dissipative force as well as it doubles the value of quadrupole force. In addition,  It can also be supported significantly as demonstrated in some recent studies via the production of the plasmonic modes near a surface \cite{lembessis2011surface,al2012generation,lembessis2009surface,al2010optical}. All these techniques could be generally described and used with ordinary light or non-ordinary one.

However, three types of light that we have investigated here can play an interesting role in the enhancement of the quadrupole interactions when we consider the higher order beams. Such kinds of high order beams have already produced in laboratories \cite{curtis2003structure,fickler2012quantum,laabs1996excitation,choporova2017high,jimenez2016formation, geneaux2017radial}. This technique, in particular, stands behind the importance of such types of light in various modern atomic applications.
The remarkable concern that can be clearly seen from these results is that quadrupole interactions, which are usually ignored in atom optics, can be taken in consideration for some crucial cases, in particular the high order beams can be produced experimentally.  It is well known that the Van der Waals interactions have a crucial feature in the adsorption of atoms and molecules on the surfaces. These characters become more considerable when the force between the adsorbate and surface is established. Due to recent experimental advances \cite{fickler2012quantum,schmiegelow2016transfer,afanasev2018experimental,sakai2018nanofocusing,hu2012v,campbell2012generation}, the quadrupole interactions can play an important role in atoms and molecules interacting with structured beam \cite{babiker2018atoms,forbes2018chiroptical}. 

\bibliography{MyBib}

\end{document}